\documentclass[useAMS,usegraphicx,usenatbib]{mn2e}

\usepackage{amsmath}
\usepackage{graphicx}
\usepackage{txfonts}
\usepackage{xcolor}
\usepackage{natbib}
\usepackage[colorlinks=true]{hyperref}
\hypersetup{allcolors={blue},allbordercolors = {red}}
\usepackage{lscape}

\title[Spectroscopic stellar parameters for giant stars]{Determination of the spectroscopic stellar parameters for 257 field giant stars\footnotemark[1]\thanks{Based on observations collected at the Paranal Observatory, ESO (Chile) with the Ultra-violet and Visible Echelle Spectrograph (UVES) of the VLT, under programmes 085.C-0062  and 086.C-0098.}}
\author[S. Alves et al.]{S. Alves$^{1,2}$\thanks{E-mail:salves@astro.puc.cl (SA); lisa.benamati@astro.up.pt  (LB)},
L. Benamati$^{3,4,5}$,
N. C. Santos$^{3,4,5}$,
V. Zh. Adibekyan$^{3,4}$, 
S. G. Sousa$^{3,4,5}$,\newauthor
G. Israelian$^{6,7}$,
J. R. De Medeiros$^{8}$,
C. Lovis$^{9}$,
S. Udry$^{9}$\\
$^1$ Instituto de Astrof\'isica, Pontificia Universidad Cat\'olica de Chile, Av. Vicu\~na Mackenna 4860, 782-0436, Macul, Santiago, Chile\\
$^2$ CAPES Foundation, Ministry of Education of Brazil, Bras\'ilia / DF, Brazil\\
$^3$ Centro de Astrof\'isica da Universidade do Porto, Rua das Estrelas, 4150-762 Porto, Portugal\\
$^4$ Instituto de Astrof\'isica e Ci\^encias do Espa\c{c}o, Universidade do Porto, Rua das Estrelas, PT4150-762 Porto, Portugal\\
$^5$ Departamento de F\'isica e Astronomia, Faculdade de Ci\^encias da Universidade do Porto, Portugal\\
$^6$ Instituto de Astrof\'isica de Canarias, 38200 La Laguna, Tenerife, Spain \\
$^7$ Departamento de Astrof\'isica, Universidad de La Laguna, 38206 La Laguna, Tenerife, Spain\\
$^8$ Departamento de F\'isica Te\'orica e Experimental, Universidade Federal do Rio Grande do Norte, Campus Universit\'ario Lagoa Nova,\\ \ \  59072-970 Natal, RN, Brasil\\
$^9$ Observatoire de Gen\`eve, Universit\'e de Gen\`eve, 51 ch. des Maillettes, 1290, Sauverny, Switzerland
}
\pagerange{\pageref{firstpage}--\pageref{lastpage}} \pubyear{2015}
\begin{document}\label{firstpage}
\date{Accepted 2015 January 26.  Received 2015 January 20; in original form 2014 November 26}
\maketitle

\begin{abstract}
The study of stellar parameters of planet-hosting stars, such as metallicity and chemical abundances, help us to understand the theory of planet formation and stellar evolution.  Here, we present a catalogue of accurate stellar atmospheric parameters and iron abundances for a sample of 257 K and G field evolved stars that are being surveyed for planets using precise radial--velocity measurements as part of the CORALIE programme to search for planets around giants. The analysis was done using a set of high--resolution and high--signal-to-noise Ultraviolet and Visible Echelle Spectrograph spectra. The stellar parameters were derived using Fe I and II ionization and excitation equilibrium methods. 
To take into account possible effects related to the choice of the lines on the derived parameters, we used three different iron line-list sets in our analysis, and the results differ among themselves by a small factor for most of stars. {For those stars with previous literature parameter estimates, we found very good agreement with our own values.} In the present catalogue we are providing new precise  spectroscopic measurements of effective temperature, surface gravity, microturbulence, and  metallicity for 190 stars for which it has not been found or published in previous articles.
\end{abstract}


\section{Introduction}

{Stellar mass and metallicity are the two main parameters that have been suggested to influence the planet formation process. To date, over 1800 extrasolar planets have been discovered and this number continues to increase. However, more than 90\% of these planets orbit main sequence stars with masses smaller than 1.50M$\odot$. Because of this, our understanding of planet formation as a function of the mass of the host star, and of the stellar environments is poorly understood.}
Precise spectroscopic studies of field dwarf stars have suggested that the frequency of giant planets is a strong function of the stellar metallicity. It seems easier to find a planet around a metal-rich star than around a metal-poor object (e.g.  \citealt*{2004A&A...415.1153S}, \citealt{2005ApJ...622.1102F}, \citealt{2011A&A...533A.141S}). This result has usually been interpreted as due to a higher probability of forming a giant planet in a metal-rich environment. Such conclusion perfectly fits the core-accretion model for giant planet formation \citep[e.g.][]{2004ApJ...616..567I,2012A&A...547A.111M}.

{Although {the} giant planet - metallicity correlation is well establish, there are number of details still missing}, and some results have cast doubts into this subject. For instance, it was proposed that the metallicity-giant planet correlation may not be present for intermediate mass stars hosting giant planets \citep{2007A&A...473..979P}. Although this conclusion is not unanimous \citep[see][]{2007A&A...475.1003H,2015A&A...574A.116R}, the question is now being debated \citep{2010ApJ...725..721G,2013A&A...554A..84M,2013A&A...551A.112M}. {Indeed, in their recent work  \citet{2015A&A...574A.116R} argued that there are consistent indications for a planet-metallicity correlation among giant star planet hosts which matches the observed planet-metallicity correlation for main-sequence hosts, in  sharp contrast with the results of \citet{2007A&A...473..979P}.} 

In fact, the discovery of several planets orbiting metal-poor objects \citep{2007ApJ...665.1407C,2008A&A...480..889S,2008A&A...487..369S} shows that giant planet formation is not completely inhibited in the metal-poor regime \citep[see also discussion in][]{2004A&A...415.1153S}.  This lends support to the idea that disc instability processes could also lead to the formation of giant planets \citep[e.g.][]{2002ApJ...567L.149B}.  {It should also be noted} that even though evolved planet hosts are {on average more metal-poor than planet-hosting dwarfs}\footnote{{However, it should be considered the possibility that {giant stellar samples that are searched for planets} may are biased.  {Hence, the comparison of dwarf stars with giant stars should be done cautiously.} }},  { there seems to be} no metallicity enhancement present for red giants with planets regarding to red giants without planets detected \citep[and references therein]{2013A&A...557A..70M}.

Concerning the intermediate mass stars hosting planets, it could be, for example, that the higher mass of these stars, competing with the stellar metallicity, is changing the observed trends. If confirmed, however, the results of \citet{2007A&A...473..979P} would cast doubts in the planet-metallicity relation observed for dwarf stars, or at least in the way this relation has been interpreted. These authors suggest that such a difference between main sequence and evolved stars is due to pollution, which is more effective for stars in the main sequence than for evolved giant stars where the convective zone enlarge and mix a large fraction of the stellar gas. Other explanations are also possible: giants are on average more massive \citep[see however][]{2013ApJ...774L...2L} than main sequence stars surveyed for planet search, therefore the frequency of planets around giants could be explained with  more sophisticated models that take into account the dependence of the snow line with the stellar mass. It may produce a large increase of giant planets frequency with the stellar mass \citep{2008ApJ...673..502K}.
In this scheme only the total amount of metals present (and not their percentage) is relevant, therefore more massive metal-poor stars (with more massive discs) may still produce many planets.

In fact, recent studies suggest that radial velocity surveys of giant stars are biased with respect to metal-poor stars, {once most of those programmes select their stellar samples with a cut-off in the $(B-V)$ colour, set to be less than or equal to 1.0. This $(B-V)$ cut-off in a sample of cool stars result in a lack of high-metallicity, low-gravity components, creating a bias that may explain} 
why planets are rather found orbiting metal-poor giants \citep{2013A&A...557A..70M}.
In addition, \citet{2009A&A...493..309S} have shown that the derivation of the parameters should be considered as another explanation, {since} the way the lines are chosen may {have} a determining role for the derivation of metallicities for giant stars.

A clear answer to these questions is fundamental for our understanding of planet formation models. One way to approach this issue would be to explore the frequency of giant planets orbiting intermediate mass stars. However, given the huge difficulties of obtaining precise radial velocities for massive main sequence stars, one of the most effective ways to access the frequency of planets around higher mass stars is to search for planets around giants. Unfortunately, it is not easy to derive the mass for a {highly evolved field} star. Red giant phase, red giant branch and horizontal branch stars with different ages, masses, and metallicities occupy a similar position in the Hertzprung--Russel (H--R) diagram, the so called mass--age--metallicity degeneracy, and therefore the mass and evolutionary status cannot be determined simply by comparing their effective temperature and luminosity with isomass tracks \citep{2011A&A...536A..71J}\footnote{{Note that using asteroseismology the degenerency observed in the H--R diagram may be broken and then we can have better accuracy/precision for mass estimation.}}. Due to this it is more difficult to study the `stellar mass-frequency of planets' relation for giant stars. A more accurate derivation of uniform and precise parameters for the giants in planet search samples is needed if we want to overcome this problem \citep{2009A&A...493..309S}.

In this paper, we present a catalogue of accurate {atmospheric} parameters for a sample of {257} field giant stars that are being surveyed for planets using precise radial-velocity measurements. These results will be very useful to study the frequency of planets as a function of the different stellar parameters, including their chemical composition ({Adibekyan} et al., in preparation) and mass,  and confront the results with model predictions. Also, homogeneous determination of parameters for comparison works are fundamental to avoid possible misconceptions{, such as systematic deviations intrinsic to the use of different approaches for measuring atmospheric parameters.}

This paper is arranged as follows. In Sect.~\ref{Sec2} we present the stellar sample. In Sect.~\ref{Sec3}  we present the method used to derive the spectroscopic parameters, and the results obtained, discussing the implications of {the} use {of} different line-lists to derive the parameters. In Sect.~\ref{Sec4}  we compare our results with literature data. We conclude in Sect.~\ref{Sec5}.

\section{Stellar Sample}\label{Sec2}

For this analysis, we used a sample of  {257} K and G evolved stars, that are being surveyed for planets using precise radial velocity measurements {in the context of the Geneva extrasolar planet search programme. {The data were obtained} at La Silla Obervatory (Chile), using the CORALIE spectrograph \citep{2000A&A...356..590U}. High-resolution ($\lambda/\Delta \lambda \sim 110000$) and high-{signal-to-noise} (S/N)  spectra for all stars in the sample were obtained using the Ultra-violet and Visible Echelle
Spectrograph(UVES) (VLT Kueyen Telescope, Paranal, Chile), between 2010 April and December (ESO programmes 085.C-0062 and 086.C-0098). The observations were done using the {UVES standard setup Red 580 (R = 47 000, spectral range: 480 -- 680  nm)}. The final spectra cover the wavelength domain between 4780 and 6805 \AA, and have a typical S/N of $\sim$ 150. All spectra for each individual star were combined using the \texttt{IRAF}\footnote{\texttt{IRAF} is distributed by National Optical Astronomy Observatories, operated by the Association of Universities for Research in Astronomy, Inc., under contract with the National Science Foundation, USA.} \texttt{SCOMBINE} task. The spectra were reduced using the UVES pipeline.
}

\begin{figure}
\centering
   \includegraphics[width=9.1cm]{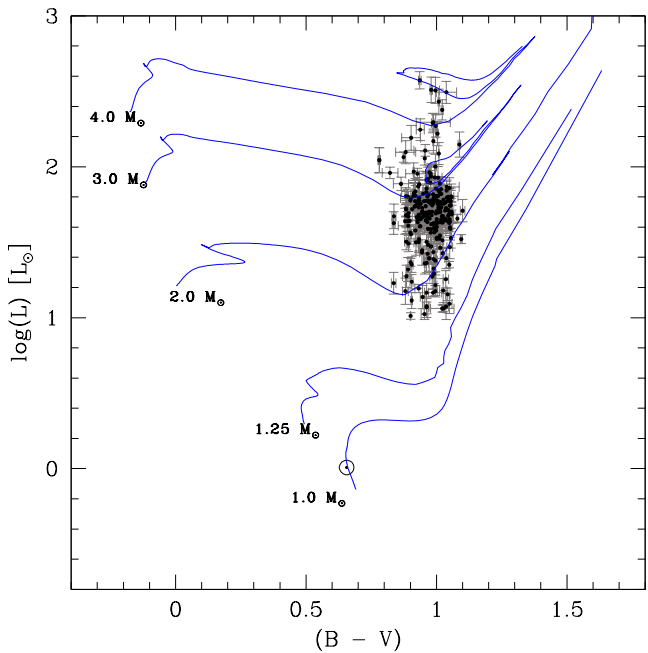}
     \caption{H--R diagram for our stellar sample. Evolutionary tracks from \citet{2012A&A...537A.146E}, for an initial abundance of metals set to $Z = 0.014$. The Sun is indicated by his usual symbol $\odot$.
}
     \label{fig1}
\end{figure}

Fig.~\ref{fig1} presents the distribution of these stars through the HR diagram. In this figure, the $(B-V)$ colour index was taken from \textit{Hipparcos} \citep{1997A&A...323L..49P}. {The surface luminosity, log $L/L\odot$, was computed 
from the estimated \textit{Hipparcos} parallaxes and $V$ magnitude following the calibrations presented by \citet{1996ApJ...469..355F}, and revisited by \citet{2010AJ....140.1158T}}. The evolutionary tracks are from \citet{2012A&A...537A.146E}, for an initial abundance of metals set to $Z$ = 0.014. This plot indicate that our sample is composed only by giant stars. As we can see in this figure, the stars in this sample have stellar masses between 1.5 and 4.0 solar masses. {For a detailed discussion on mass distribution of (sub-)giant planet hosts we redirect the reader to \citet{2013ApJ...774L...2L}.}

\section{Determination of spectroscopic parameters}\label{Sec3}

The atmospheric stellar parameters -- the effective temperature $T_{\textrm{eff}}$, the surface gravity $\log g$, the microturbulence $\xi$, and the metallicity [Fe/H] -- were derived following the same procedure described in \citet{2004A&A...415.1153S}. Such a procedure is based on the equivalent widths (EW) of Fe I and Fe II lines, and iron excitation and ionization equilibrium, assumed a local thermodynamic equilibrium. For this mostly automatic analysis we used the code \texttt{MOOG}\footnote{\url{http://www.as.utexas.edu/~chris/moog.html}} \citep{1973PhDT.......180S}, with a grid of ATLAS plane--parallel model atmospheres \citep{1993KurCD..13.....K}. The EWs were computed by using the code \texttt{ARES} \citep[Automatic Routine for line Equivalent widths in stellar spectra ---][]{2007A&A...469..783S}\footnote{{For a general review of the \texttt{ARES+MOOG} method used here see \citet{2014arXiv1407.5817S} }}. The input parameters for \texttt{ARES} are the same as in \citet{2008A&A...487..373S}, and the S/N adopted is given as a root sum square of the S/N of the order related to 6000~\AA\ (see UVES ETC)\footnote{\url{http://www.eso.org/observing/etc/bin/gen/form?INS.NAME=UVES+INS.MODE=spectro}}, for each spectrum, extracted in the header of the fits reduced spectra. The determination of the uncertainties in the derived parameters also follow the same prescription as in \citet{2004A&A...415.1153S}.

{The determination of spectroscopic stellar parameters greatly depends on the selection of the Fe lines and the transition probabilities of those lines.} For a complete discussion about this dependence and its implications for the study of chemical abundances in giant stars, we direct the reader to \citet{2009A&A...493..309S}. These authors also discuss the implications of a carefully choice of the lines for derivation of metallicities and other stellar parameters. Due to the fact that stellar parameters may depend on the line-list choice, for our analysis we use three different line-lists: the large line-list from \citet[][hereafter SO08]{2008A&A...487..373S}, the line-list for cooler stars based on the SO08 line-list \citep[][hereafter TS13]{2013A&A...555A.150T}, and the small line-list, made specifically for giant stars, from \citet[][hereafter HM07]{2007A&A...475.1003H}.
{SO08 is composed by 263 Fe I abd 36 Fe II weak lines, and the transition probabilities ($\log gf$ values) were determined using a differential analysis to the Sun. The reference solar iron abundance used to make this list is $A$(Fe)$\odot$ = 7.47. TS13 was built from this list, specifically for cooler stars. As only weak and isolated lines were left, to avoid blending effects, the TS13 line-list is composed by 120 Fe I and 17 Fe II lines. The smaller line-list is HM07, with 16 Fe I and 6 Fe II lines. This line-list was specifically made for giant stars and all lines were carefully selected to avoid blends by atomic and CN lines. The $\log gf$ values are mostly based on different laboratory works (some few cases with small adjustments using the Arcturus atlas), as there is no single source of laboratory $\log gf$ for either Fe I or Fe II. Unlike the SO08 and TS13 line-lists, the reference value of $A$(Fe)$\odot$ = 7.49 is used.}
Table~\ref{Tbl1} lists all stellar parameters derived with the SO08, TS13 and HM07 line-lists. Note that no viable solution could be found for HD~74006 using the HM07 line-list.

As pointed out by TS13 the results from using the SO08 line-list for cool stars ($T_{\textrm{eff}} < $ 5200K) have shown to be unsatisfactory. For those stars, the results found with the TS13 line-list presents a better agreement with the expected value \citep[for a detailed description of this see][]{2013A&A...557A..70M}. Due to the fact that most of stars in our sample have temperatures lower than 5200 K, we adopt as final the parameters derived using the TS13 line-list. For the stars with temperatures greater than 5200 K, {one} still can adopt the results of the SO08 line-list. These stars are highlighted in Table~\ref{Tbl1}. {Please note that in the present analysis we use only the parameters derived using the TS13 line-list. In addition, particular attention should be taken when using the results presented in Table~\ref{Tbl1} concerning the possibility to round the numbers, especially for effective temperature and microturbulence. We have chosen not to round the numbers, but present them as they were delivered by the code.}

\begin{figure*}
\centering
    \includegraphics[width=17cm]{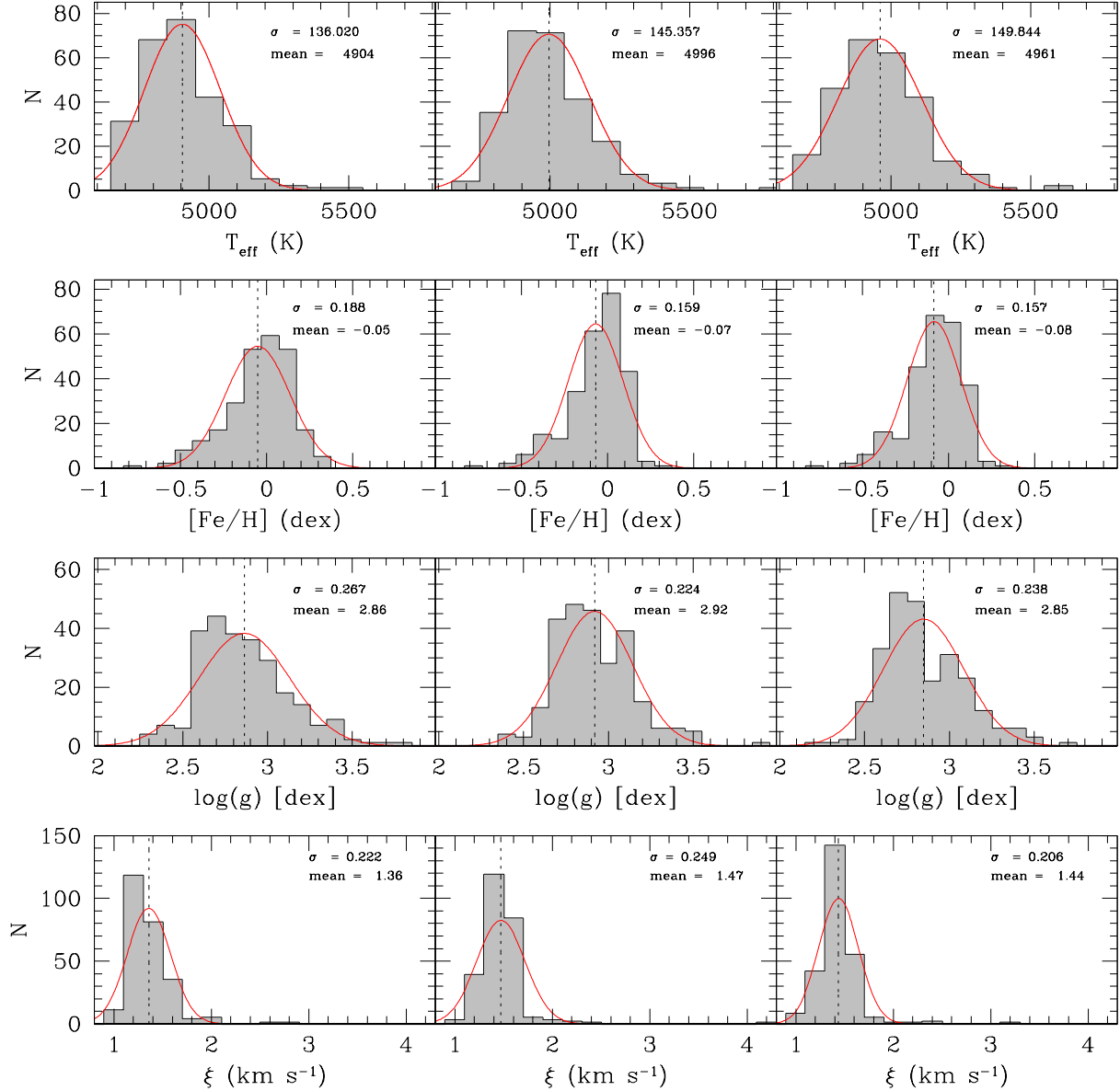}
     \caption{Distribution of the atmospheric parameters derived using the HM07 (left-hand panel), SO08 ( middle panel), and TS13 (right-hand panel) line-lists. In each plot, the Gaussian fit to the distribution, along with the values of the mean ({dotted line}) and the standard deviation $\sigma$ is also shown.}
     \label{Fig2}
\end{figure*}

\begin{figure*}
\centering
    \includegraphics[width=19cm]{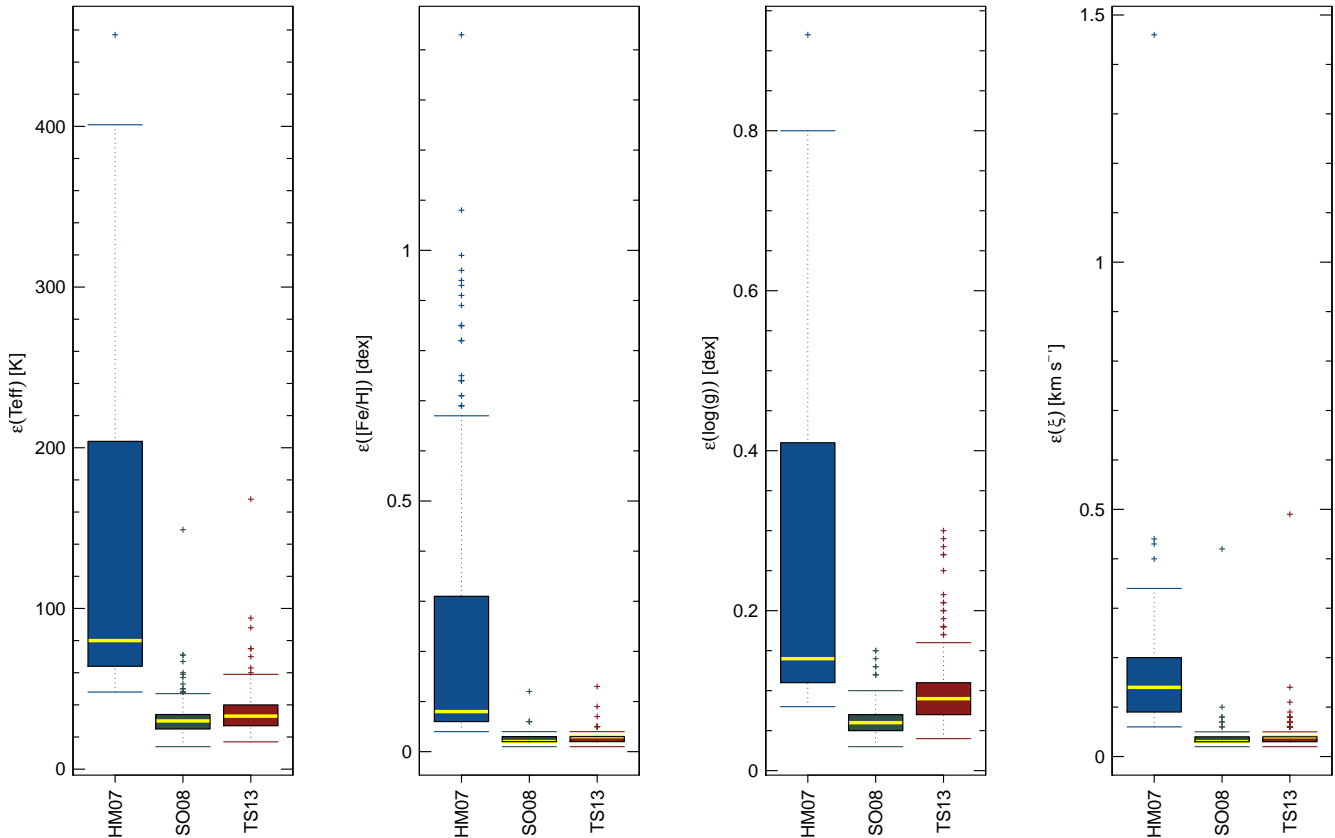}
     \caption{
     Boxplot showing the median (solid horizontal yellow lines), lower and upper quartiles (box), range of data points within 1.5$\times$ (75 -- 25 per cent) range (whiskers), and outliers (individual crosses) of the uncertainties for derived parameters presented in Fig.~\ref{Fig2}.}
     \label{Fig2b}
\end{figure*}

In Fig.~\ref{Fig2}, we present the distribution of the atmospheric stellar parameters derived in our work, using, respectively, the HM07 (left-hand panel), SO08 (middle panel), and TS13 (right-hand panel) line-lists. We can see a general agreement for the results found for the three cases (HM07, SO08, and TS13 line-lists).  The results from TS13 and SO08 are compatible in terms of metallicity, but in effective temperature they show an offset, specially for the cooler stars, as expected and discussed in TS13 and \citet{2013A&A...551A.112M}. The uncertainties are illustrated in the boxplot in Fig.~\ref{Fig2b} which shows the median and quartiles for each parameters derived with the HM07 ({left-hand panel}), SO08 ({middle panel}), and TS13 ({right-hand panel}) line-lists. As we can see in this figure, HM07 results show a much higher dispersion on the uncertainties, as expected given their smaller number of lines. These relatively small number of lines (16 Fe I and 6 Fe II) decrease the statistical strength of the determined stellar parameters, increasing the internal dispersion, especially for Fe II. For instance, the metallicity derived using this line-list present high values for the errors, with an average of about 0.22~dex, and $\epsilon$ [Fe/H] $>$ 1.0 dex for nine stars.

\begin{figure*}
\begin{minipage}{165mm}
\begin{minipage}[b]{0.45\linewidth}
\includegraphics[width=8cm]{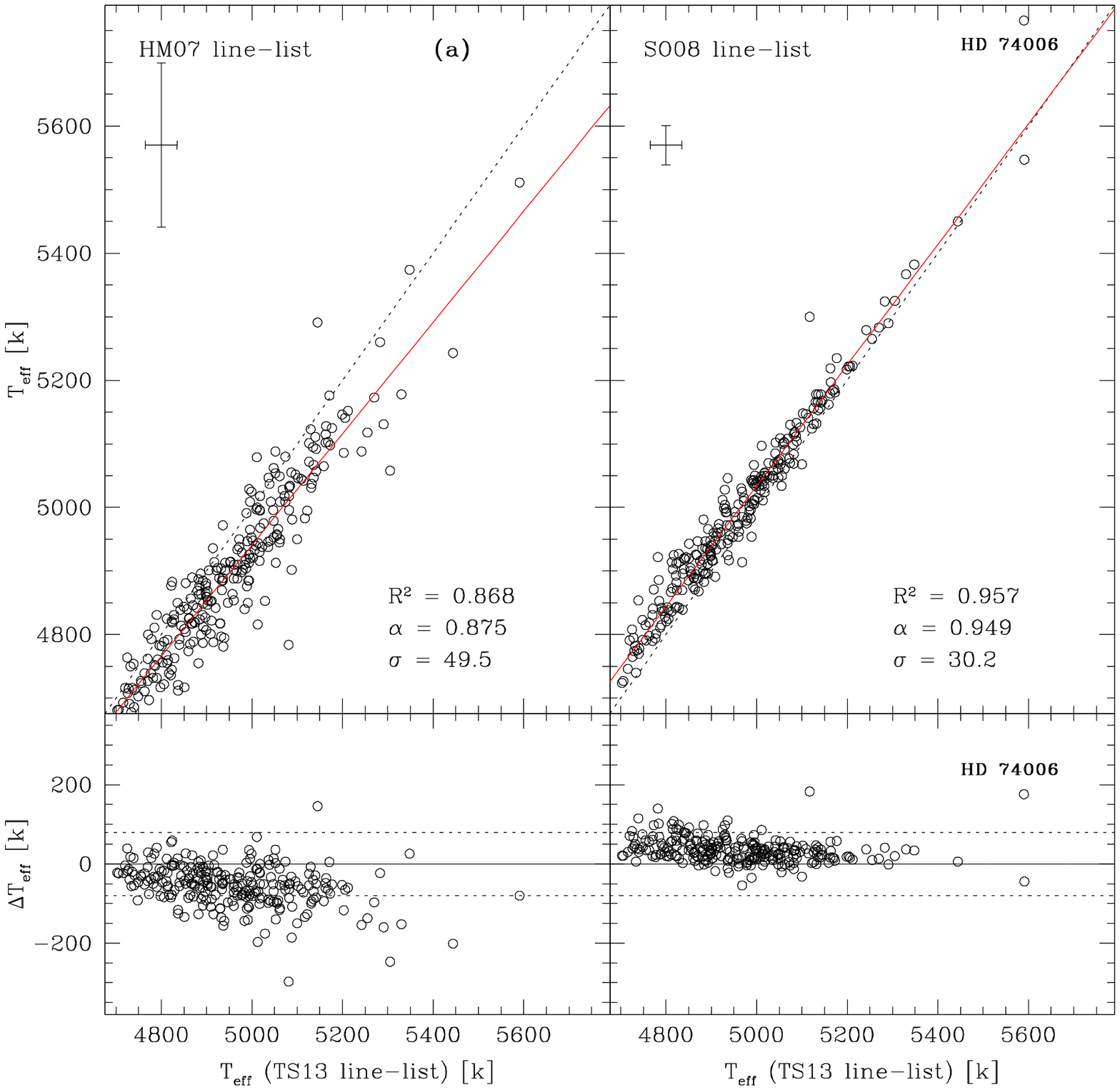}
\end{minipage}\hfill
\begin{minipage}[b]{0.45\linewidth}
\includegraphics[width=8cm]{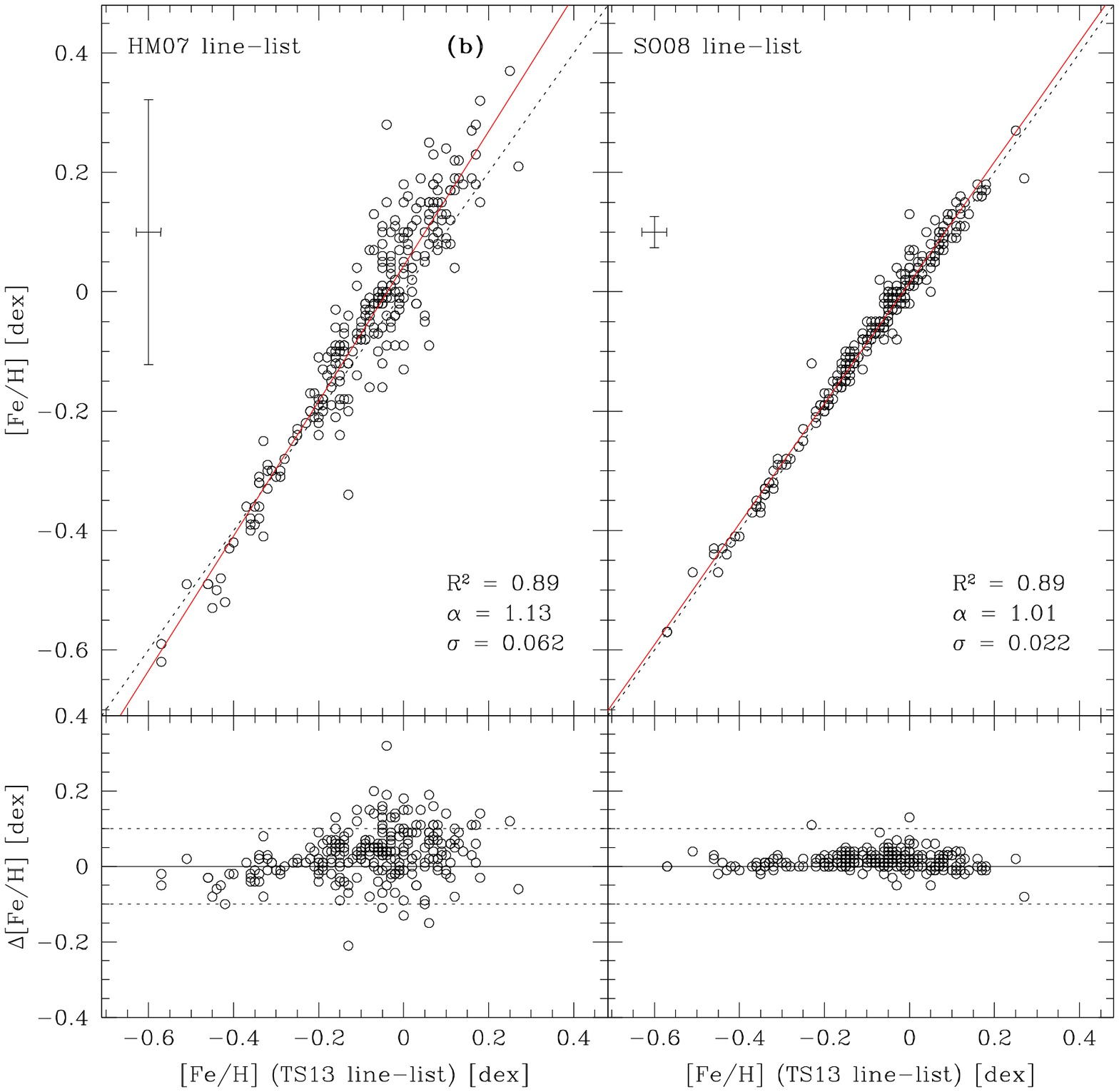}
\end{minipage}
\begin{minipage}[b]{0.45\linewidth}
\includegraphics[width=8cm]{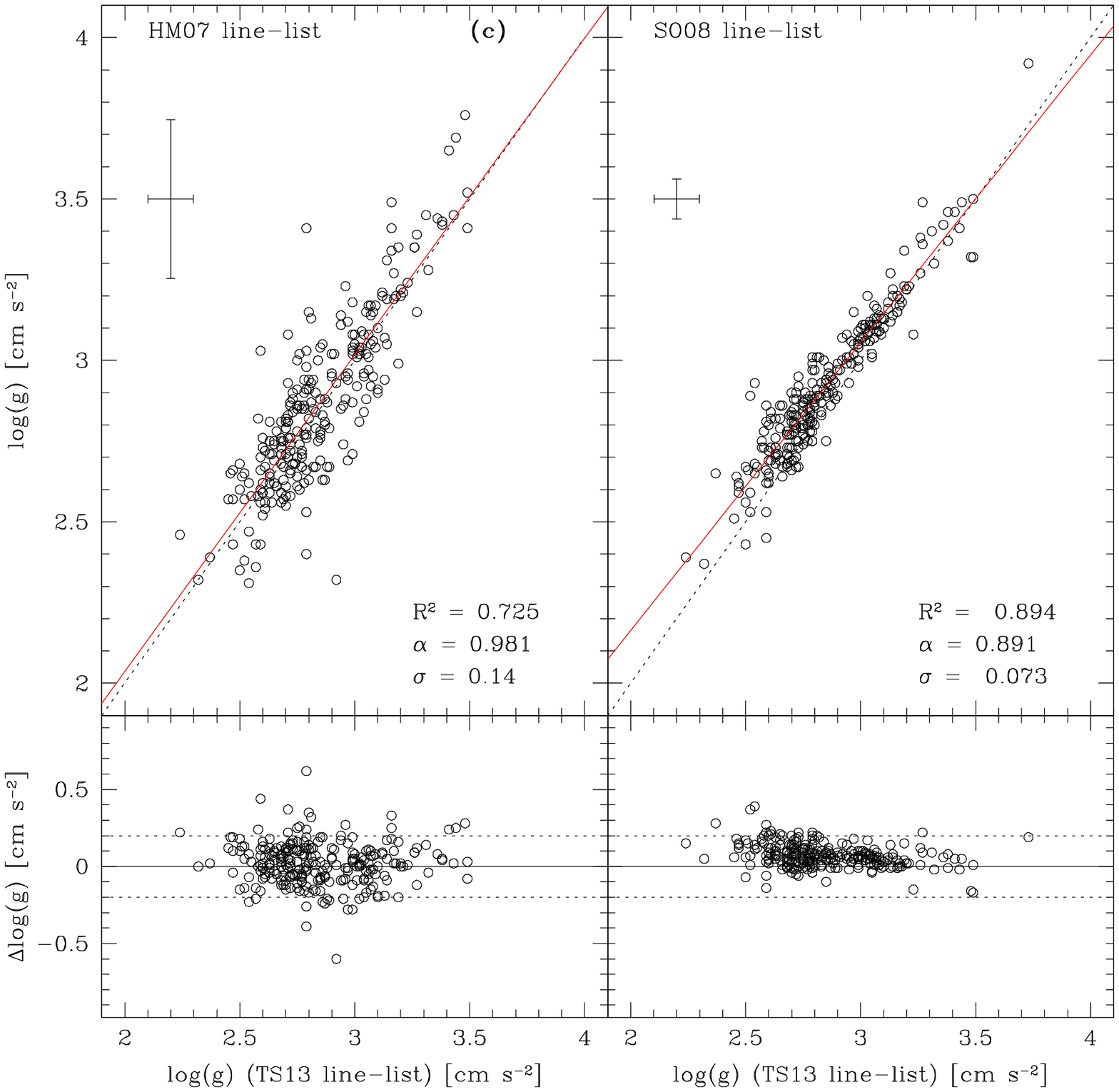}
\end{minipage}\hfill
\begin{minipage}[b]{0.45\linewidth}
\includegraphics[width=8cm]{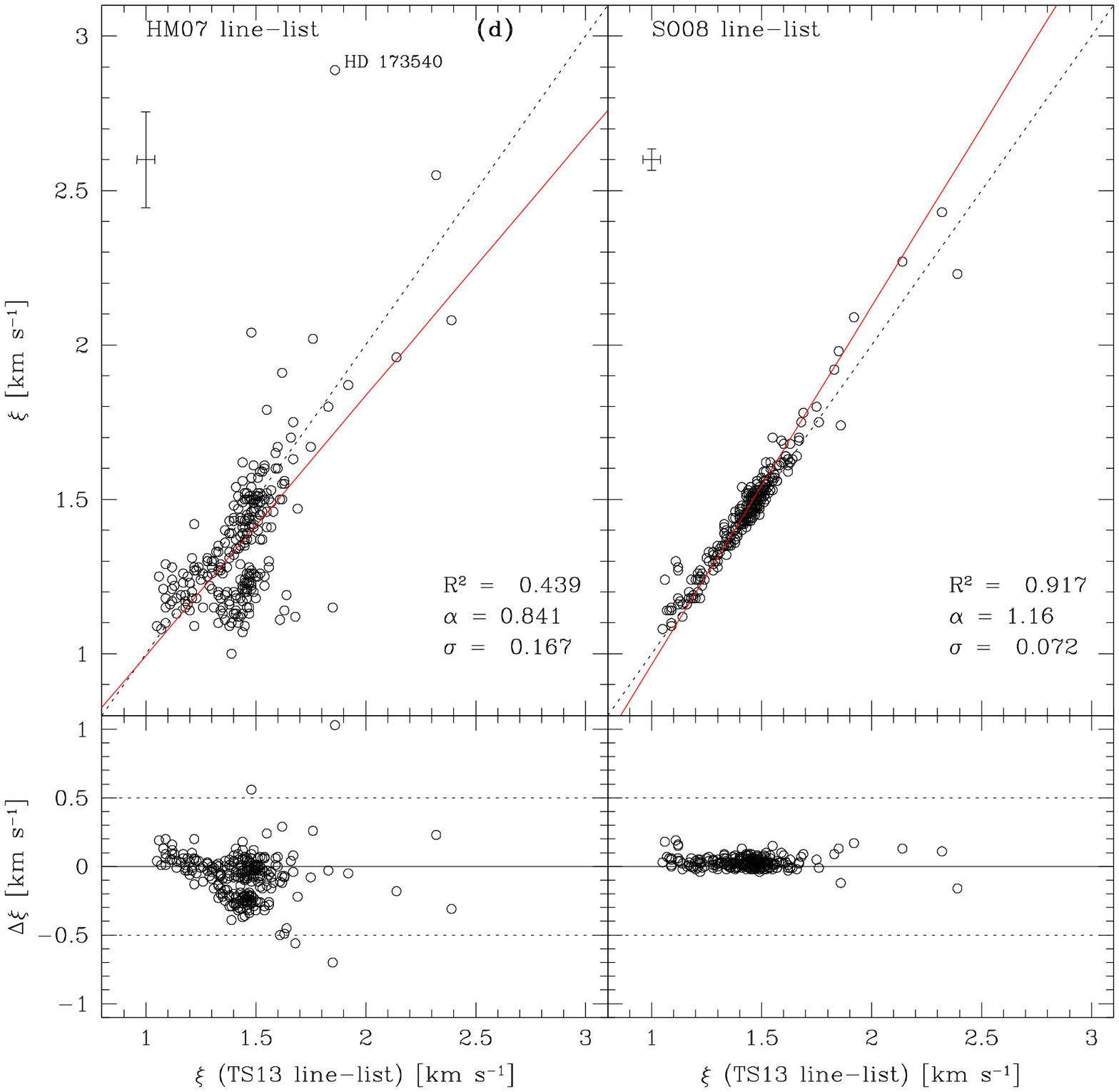} 
\end{minipage}
\caption{Comparison between the results found using the TS13, SO08, and HM07 line-lists. \textbf{(a)} Effective temperature, \textbf{(b)} metallicity, \textbf{(c)} surface gravity, and \textbf{(d)} microturbulence velocity derived using the HM07 ({left-hand plot}) and SO08 ({right-hand plot}) line-lists compared to those derived using the TS13 line-list. In each panel, the lower plot compares the differences from perfect agreement. The differences refer to the  abscissa minus the ordinate  of the corresponding upper plot. The dotted line shows the one-to-one relation, and the solid line is the linear fit, for which is given the values of the  $R$-squared $R^2$, the slope calculated by the regression $\alpha$,  and the residual standard deviation $\sigma$.  The average error is also plotted. HD 74006 is not shown in the plots presenting the results found using the HM07 line-list.}
     \label{Fig3}
\end{minipage}
\end{figure*}

Fig.~\ref{Fig3} shows the comparison between the results obtained using the HM07 and SO08 line-lists with those obtained using the TS13 line-list which we adopt as final parameters. {As pointed out before, we adopt the TS13 results because this line-list was built specifically for stars with $T_\textrm{{eff}} < 5200 $K. Even for some stars that have higher $T_\textrm{{eff}}$, we still adopt the values derived using the TS13 line-list, since these results are in perfect agreement with those derived using SO08, within the errors. If the results of SO08 and TS13 were not in such a remarkable agreement, we would adopt the results obtained using SO08, since this is a better line-list for the hottest stars (more line hence better precision and lower error).}  Between the results of TS13 and HM07, we found an average difference  (defined as TS13-HM07 results) of  54 K, -0.019 dex, 0.075 km s$^{-1}$, and -0.032 dex for effective temperature, surface gravity, microturbulence, and metallicity, while between the results of TS13 and SO08 the average differences (defined as TS13-SO08 results) are -35 K, -0.071 dex, -0.033 km s$^{-1}$, and -0.015~dex. {Such good agreement between the results of SO08 and TS13 may be due to the good quality of the spectra, with both high resolution and high S/N, in which case the SO08 line-list is probably less affected by blended lines.} Microturbulences compares very well with SO08 line-list, but the results found with HM07 line-list are slightly higher than the other values, but still within the error bar. {As a matter of fact, microturbulence is the most affected parameter when the smaller line-list of HM07 was adopted. In the bottom right of Fig.~\ref{Fig3}, panel (d), we can see that a group of stars is separated from the others in the trend, and thus for these stars the HM07 microturbulence results differ significantly from those obtained with the TS13 line-list. We may explain those large discrepancies due to the small EW interval of Fe I lines measured using the  HM07 line-list, which did not allow a consistent determination of the microturbulence. In addition, note that HD~173540 has an abnormal error, $\xi =  2.89 \pm 1.46$ km s$^{-1}$. If, instead, we use the empirical formula derived by HM07 based on their results, $\xi = 3.7 - 5.1 \times 10^{-4} T_\textrm{{eff}}$, to estimated the microturbulence for this star, we get $\xi = 0.9 \pm 0.06$ km s$^{-1}$. Furthermore, a more reliable value ($\xi = 1.2 \pm 0.01 $ km s$^{-1}$) was found applying the formula from \citet{2013ApJ...764...78R}, $\xi = 1.163 + 7.808 \times 10^{-4}(T_{eff} - 5800) - 0.494(\log g - 4.30) - 0.050$ [Fe/H]}. 
{A more deeper and detailed discussion about these offsets are presented by \citet{2013A&A...557A..70M}.}

\section{Comparison with previous works} \label{Sec4}

\begin{figure*}
\centering
  \includegraphics[width=17cm]{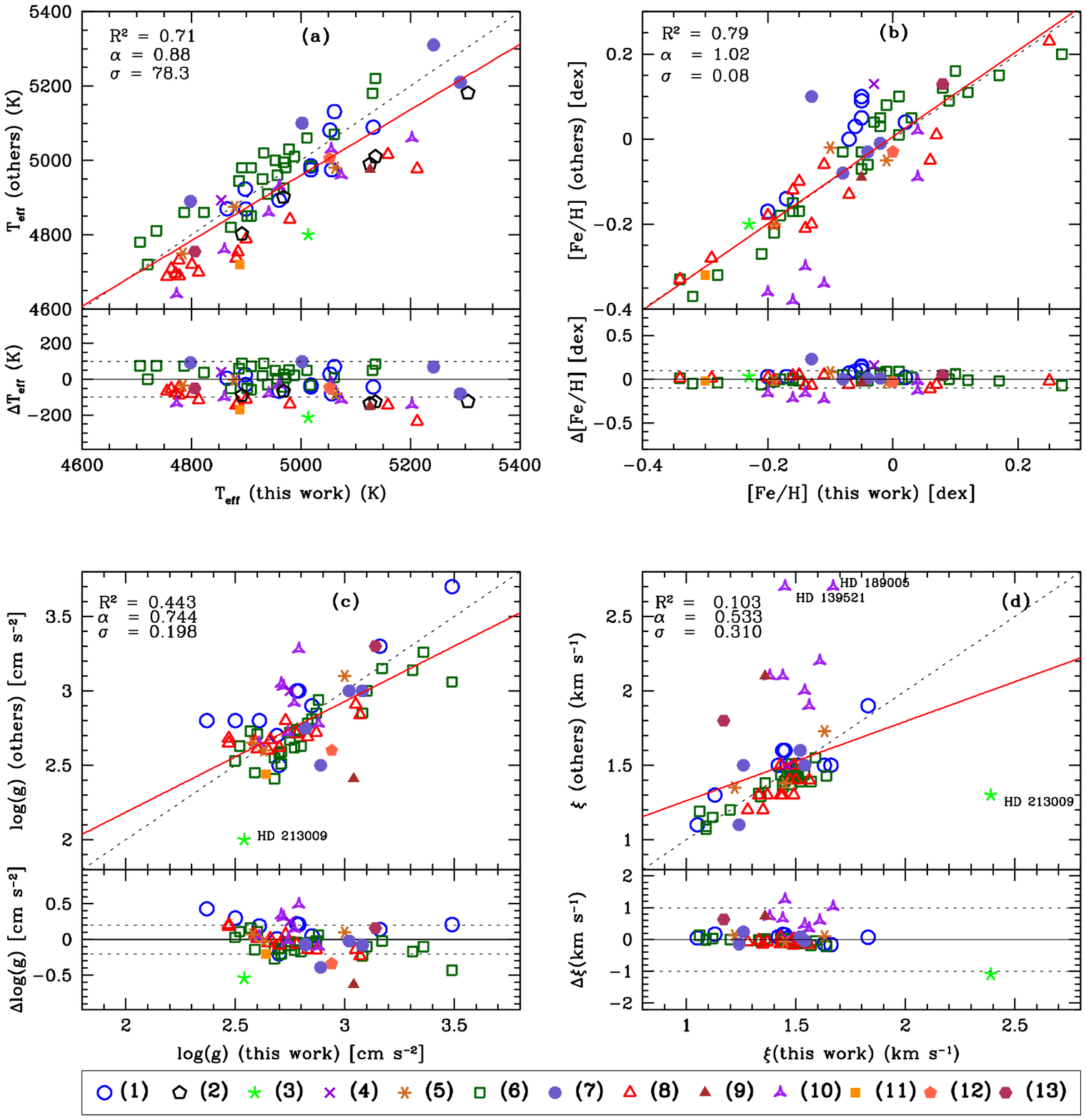}
     \caption{Comparison of the results of this work obtained for the TS13 line-list with available literature data for \textbf{(a)} effective temperature, \textbf{(b)} metallicity, \textbf{(c)} surface gravity, and \textbf{(d)} microturbulence. The dotted line shows the one-to-one relation, and the solid line is the linear fit, for which is given the values of the  $R$-squared $R^2$, the slope calculated by the regression $\alpha$,  and the residual standard deviation $\sigma$. Each symbol indicates a reference given in Table~\ref{tbl2}, as enumeration given in the legend, i. e., \textbf{(1)}:  \citet{2006A&A...458..609D}; 
     \textbf{(2)}:  \citet{1998A&A...339..858D}; 
     \textbf{(3)}:  \citet{1981A&A....93..315F}; 
     \textbf{(4)}:  \citet{1986A&A...169..201G}; 
     \textbf{(5)}:  \citet{2007A&A...475.1003H}; 
     \textbf{(6)}:  \citet{2011A&A...536A..71J}; 
     \textbf{(7)}:  \citet{1992MNRAS.256..535J}; 
     \textbf{(8)}:  \citet{2007MNRAS.382..553L}; 
     \textbf{(9)}:  \citet{1991ApJS...75..579L}; 
     \textbf{(10)}:  \citet{1990ApJS...74.1075M}; 
     \textbf{(11)}:  \citet{2008A&A...484L..21M}; 
     \textbf{(12)}:  \citet{1999A&A...348..487R}; 
     \textbf{(13)}:  \citet{2004A&A...425..187T}.}
     \label{Fig4}
\end{figure*}

The large majority of the giant stars studied in this paper do not have any previous metallicity estimate derived from high-resolution spectroscopy. In order to compare our results with previous ones, we used several works (\citealt{1981A&A....93..315F,1986A&A...169..201G,1990ApJS...74.1075M,1991ApJS...75..579L,1992MNRAS.256..535J,1998A&A...339..858D,1999A&A...348..487R}, \citealt*{2004A&A...425..187T}, \citealt {2006A&A...458..609D,2007A&A...475.1003H,2007MNRAS.382..553L,2008A&A...484L..21M,2010A&A...515A.111S,2011yCat..35369071J,2011A&A...536A..71J}) to compile a list of literature data for a set of 74 stars in our sample. The literature values of the atmospheric parameters for these common samples are listed in Table~\ref{tbl2}. Note that only 67 of these stars have all four parameters already calculated in previous works (see the last rows of the Table~\ref{tbl2}), hence we are providing here new precise spectroscopic atmospheric parameters for {an amount of} 190 stars. Fig.~\ref{Fig4} show the comparison between our results obtained for the TS13 line-list, with those presented in these earlier works. As we can see in the panels of this figure, our results present good agreement with those listed in the literature.

The atmospheric parameters taken from literature for the 74 stars presented in the Table~\ref{tbl2} can also be found in the PASTEL catalogue \citep{2010A&A...515A.111S} but not the microturbulence velocity.  Compared to the PASTEL catalogue we found an average difference  (defined as TS13 -- literature data) of 108 K, -0.02 dex, and 0.03 dex for effective temperature, surface gravity, and metallicity, respectively. 

\begin{figure*}
\centering
  \includegraphics[width=17cm]{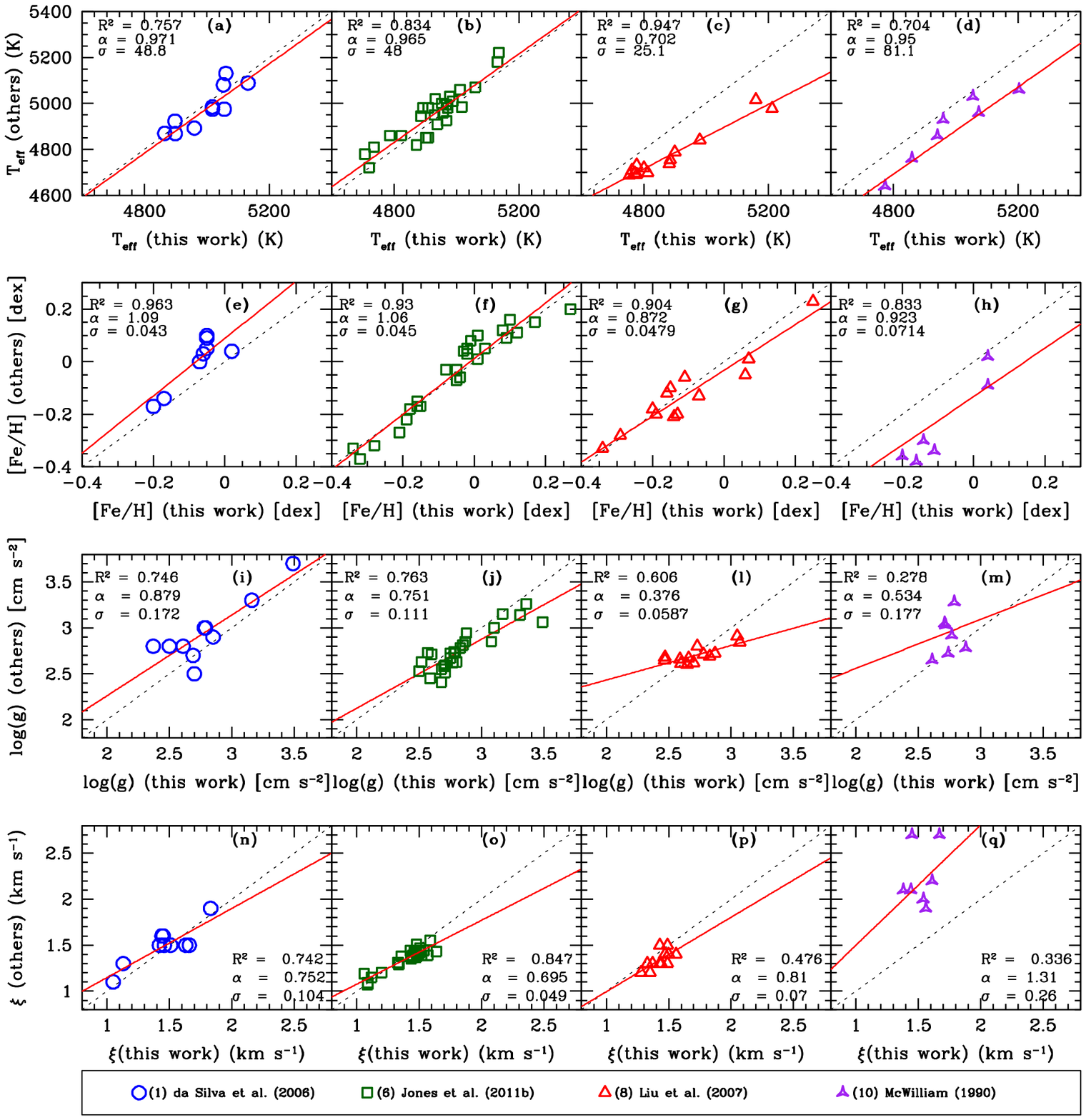}
     \caption{Comparison between our results with those from \citet{2006A&A...458..609D},   \citet{2011A&A...536A..71J}; \citet{2007MNRAS.382..553L}, and  \citet{1990ApJS...74.1075M}. The dotted line shows the one-to-one relation, and the solid line is the linear fit, for which is given the values of the  R-squared $R^2$, the slope calculated by the regression $\alpha$,  and the residual standard deviation $\sigma$. Each symbol indicates a reference given in Table~\ref{tbl2},  as enumeration given in the legend.
}
     \label{Fig5}
\end{figure*}

The common sample presented in Table~\ref{tbl2} is composed by values taken from 13 different works. In order to test our results against samples homogeneously characterized,  we checked our results,  separately,  against those from \citet{2011A&A...536A..71J}, \citet{2007MNRAS.382..553L}, \citet{1990ApJS...74.1075M}, and \citet{2006A&A...458..609D} due to the significant number of stars in common with these works. Fig.~\ref{Fig5} shows the comparison of our stellar parameters with those from these works. We found an average difference of 20 K, -0.17 dex, -0.032 km s$^{-1}$, and -0.072 dex, respectively, for effective temperature, surface gravity, microturbulence, and metallicity when we compare our results with those from \citet{2006A&A...458..609D}, and -26 K, 0.085 dex, 0.048 km s$^{-1}$, and -0.0088 dex compared to \citet{2011A&A...536A..71J}. 

We have 20 stars in common with \citet{1990ApJS...74.1075M}, {who analysed 671 GK giant spectra, and derived effective temperatures with empirical and semi-empirical methods, involving an IR flux calibration}. For this set of stars, the average difference on effective temperature is 119 K, with a standard deviation of 81.1 K, and it is less than 0.12 dex  in metallicity, with a standard deviation of 0.07 dex.  Besides the effective temperature and the metallicity from \citet{1990ApJS...74.1075M} are marginally higher than the one derived in our work, the two other parameters compare quite well, with an average difference of -0.21 dex and -0.65 km s$^{-1}$ for surface gravity and microturbulence, respectively. 

The comparison of our atmospheric parameters with those from \citet{2007MNRAS.382..553L}, with whom we have 14 stars in common, is also presented in Fig.~\ref{Fig5}. The average differences are 108 K, 0.017 dex, 0.016 dex, and 0.089 km s$^{-1}$, respectively,  for effective temperature, metallicity, surface gravity, and microturbulence. 

\nocite{*}

\section{Concluding Remarks}\label{Sec5}

In this work, we have derived the stellar atmospheric parameters (the effective temperature, the surface gravity, the microturbulence, and the metallicity) for a sample of 257 field giant stars that are being surveyed for planets using precise radial--velocity measurements. Those parameters were derived by using three different line-lists of Fe I and Fe II (SO08, TS13, and HM07). All parameters derived in this work are listed in Table~\ref{Tbl1}, and we adopt as final the parameters derived with the TS13 line-list.  When one compares the results found by using the different line-lists we found small dispersion for most of the stars.

In the present catalogue  (Table~\ref{Tbl1}), we are providing new precise  spectroscopic measurements of atmospheric parameters for 190 stars for which  the given four parameters had not yet been found or published in previous articles. Additionally, we also provide new measurements for 67 stars with previous published results of all parameters, but with the major advantage that they are now calculated homogeneously, providing a more suitable analysis. The comparison of our results with those presented in the literature shows that our derivations are solid, and it will be very useful to future studies of frequency of planets as a function of the different stellar parameters, as a comparison sample.

Since the first discovery of a substellar companion orbiting a giant star (HD 137759 -- \citealt{2002ApJ...576..478F}), more than 100 evolved stars are known to host planets according to the available data at the Extrasolar Planets Encyclopaedia\footnote{\url{http://exoplanet.eu}}, but it is still missing a homogeneous sample that allows us to perform studies on the properties of giant stars hosting planets. Note that one star in our sample is already known to have an orbiting planet (HD~11977 -- \citealt{2005A&A...437L..31S}). The parameters for this star are $T_\textrm{{eff}}$ = 5018 $\pm$ 27 K, $\log g$ = 2.85 $\pm$ 0.07 cm s$^{-2}$, $\xi$= 1.44	$\pm$ 0.03 km s$^{-1}$,  [Fe/H] = -0.17$\pm$ 0.03 dex (TS13 line-list), showing that its metal content is a bit less than that of the Sun. Low metal abundance has been also found in other giants hosting planets suggesting that planet-hosting giant stars are on average metal-poor than planet-hosting dwarfs. However, as pointed out by \citet{2013A&A...551A.112M}, it may be due to a bias in samples of evolved stars used to detect planets.  For a more complete discussion of this subject the reader is directed to \citet{2013A&A...557A..70M} and \citet*{2013A&A...554A..84M}.

In the present catalogue, the red giant branch star HD 135760 is the most metal-rich ([Fe/H] = +0.27 $\pm$ 0.05 dex), which is in agreement with previous result presented by  \citet{2011A&A...536A..71J}, while HD 7082 is the most metal--poor ([Fe/H] = -0.74 $\pm$ 0.02 dex), amongst with three other stars that have [Fe/H] $<$ -0.5 dex. 

{One of the major advantage of our work is to present a homogeneous calculation of spectroscopic parameters for a set of giant stars that have been already surveyed for planets, thus presenting a solid sample of comparison for future researches. Indeed,} most stars of our sample have already a large number of measurements of precise radial velocities with the CORALIE spectrograph spread  over the last years. Once a significant sample of planets is found in the present sample, we will be able to do analysis of the planet frequency as a function of metallicity and stellar mass. Until then, we can use our accurate and uniform stellar parameters as control sample to others studies that compares stars hosting planets with stars without detected planets. In addition, a complete study of chemical abundances for those stars (i. e., Table~\ref{Tbl1}) will be released by Adibekyan et al. (in preparation).

\section*{Acknowledgements}
This work was supported by the European Research Council/European Community under the FP7 through Starting Grant agreement number 239953. This work was also supported by the Gaia Research for European Astronomy Training (GREATITN) Marie Curie network, funded through the European Union Seventh Framework Programme ([FP7/2007-2013]) under grant agreement number 264895. NCS was supported by FCT through the Investigator FCT contract reference IF/00169/2012 and POPH/FSE (EC) by FEDER funding through the programme `Programa Operacional de Factores de Competitividade' - COMPETE.
VZhA and SGS are supported by grants SFRH/BPD/70574/2010 and SFRH/BPD/47611/2008 from the FCT (Portugal), respectively.
Research activities of the Obsevational Stellar Board of the Federal University of Rio Grande do Norte are supported by continuous grants of CNPq and FAPERN brazilian agencies and by the INCT-INEspa\c{c}o.
SA acknowledges Post-Doctoral Fellowship from the CAPES brazilian agency  (PNPD/2011: \textit{Concess\~ao Institucional}), Post-Doctoral Fellowship CAPES/PDE (BEX-2077140), and also support by Iniciativa Cient\'ifica Milenio through grant IC120009, awarded to The Millennium Institute of Astrophysics. {Authors are very thankful to referee, Chris Sneden, for his very constructive and helpful remarks that helped us to improve the paper.}\\


%
\bibliographystyle{mn2e}
\bibliography{refs} 
\bsp

\begin{landscape}
\begin{table}
\caption{Stellar parameters determined from the iron lines by using HM07, SO08, and TS13 line-lists. The adopted parameters in our work is the one derived with the TS13 line-list.\label{Tbl1}}
\begin{tabular}{r|lccc|lccc|lccc}
\hline\hline
 & \multicolumn{4}{c|}{HM07 line-list}	& \multicolumn{4}{c|}{S08 line-list} 	& \multicolumn{4}{c}{T13 line-list}	\\	\hline		
HD	&	T$_{eff}$			&	$\log g$			&	$\xi$			&	[Fe/H]			&	T$_{eff}$			&	$\log g$			&	$\xi$			&	[Fe/H]			&	T$_{eff}$			&	$\log g$			&	$\xi$			&	[Fe/H]			\\
number	&	(K)			&	(cm s$^{-2}$)			&	(km s$^{-1}$)			&	(dex)			&	(K)			&	(cm s$^{-2}$)			&	(km s$^{-1}$)			&	(dex)			&	(K)			&	(cm s$^{-2}$)			&	(km s$^{-1}$)			&	(dex)			\\
\hline
496	&	4881	$\pm$	225	&	2.78	$\pm$	0.45	&	1.21	$\pm$	0.20	&	0.05	$\pm$	0.18	&	4900	$\pm$	39	&	2.79	$\pm$	0.07	&	1.52	$\pm$	0.04	&	-0.03	$\pm$	0.03	&	4854	$\pm$	40	&	2.75	$\pm$	0.10	&	1.46	$\pm$	0.04	&	-0.03	$\pm$	0.03	\\
636	&	4784	$\pm$	202	&	2.86	$\pm$	0.41	&	1.26	$\pm$	0.20	&	0.24	$\pm$	0.31	&	4903	$\pm$	46	&	2.87	$\pm$	0.09	&	1.50	$\pm$	0.04	&	0.13	$\pm$	0.03	&	4840	$\pm$	43	&	2.75	$\pm$	0.11	&	1.46	$\pm$	0.04	&	0.10	$\pm$	0.03	\\
770	&	4800	$\pm$	208	&	2.66	$\pm$	0.42	&	1.25	$\pm$	0.20	&	-0.03	$\pm$	0.32	&	4841	$\pm$	35	&	2.59	$\pm$	0.07	&	1.52	$\pm$	0.03	&	-0.13	$\pm$	0.03	&	4771	$\pm$	45	&	2.47	$\pm$	0.10	&	1.50	$\pm$	0.04	&	-0.16	$\pm$	0.03	\\
1737	&	4797	$\pm$	203	&	2.36	$\pm$	0.42	&	1.25	$\pm$	0.20	&	0.13	$\pm$	0.51	&	4998	$\pm$	47	&	2.75	$\pm$	0.08	&	1.57	$\pm$	0.05	&	0.13	$\pm$	0.04	&	4932	$\pm$	50	&	2.57	$\pm$	0.12	&	1.54	$\pm$	0.05	&	0.10	$\pm$	0.04	\\
3488	&	4869	$\pm$	63	&	2.87	$\pm$	0.11	&	1.56	$\pm$	0.09	&	-0.08	$\pm$	0.06	&	4935	$\pm$	30	&	2.83	$\pm$	0.06	&	1.50	$\pm$	0.03	&	-0.08	$\pm$	0.03	&	4880	$\pm$	34	&	2.72	$\pm$	0.08	&	1.44	$\pm$	0.04	&	-0.10	$\pm$	0.03	\\
\end{tabular}
\end{table}
\end{landscape}
\begin{landscape}
\begin{table}
\contcaption{}
\begin{tabular}{r|lccc|lccc|lccc}
\hline\hline
 & \multicolumn{4}{c|}{HM07 line-list}	& \multicolumn{4}{c|}{S08 line-list} 	& \multicolumn{4}{c}{T13 line-list}	\\	\hline		
HD	&	T$_{eff}$			&	$\log g$			&	$\xi$			&	[Fe/H]			&	T$_{eff}$			&	$\log g$			&	$\xi$			&	[Fe/H]			&	T$_{eff}$			&	$\log g$			&	$\xi$			&	[Fe/H]			\\
number	&	(K)			&	(cm s$^{-2}$)			&	(km s$^{-1}$)			&	(dex)			&	(K)			&	(cm s$^{-2}$)			&	(km s$^{-1}$)			&	(dex)			&	(K)			&	(cm s$^{-2}$)			&	(km s$^{-1}$)			&	(dex)			\\
\hline
22366	&	4678	$\pm$	65	&	2.68	$\pm$	0.12	&	1.40	$\pm$	0.08	&	-0.36	$\pm$	0.05	&	4740	$\pm$	22	&	2.67	$\pm$	0.04	&	1.35	$\pm$	0.02	&	-0.37	$\pm$	0.02	&	4734	$\pm$	27	&	2.68	$\pm$	0.06	&	1.36	$\pm$	0.03	&	-0.37	$\pm$	0.02	\\
22382	&	4809	$\pm$	215	&	2.72	$\pm$	0.43	&	1.25	$\pm$	0.20	&	0.10	$\pm$	0.53	&	4920	$\pm$	35	&	2.90	$\pm$	0.07	&	1.49	$\pm$	0.03	&	0.07	$\pm$	0.03	&	4851	$\pm$	38	&	2.72	$\pm$	0.09	&	1.51	$\pm$	0.04	&	0.01	$\pm$	0.03	\\
22532	&	4996	$\pm$	73	&	3.22	$\pm$	0.12	&	1.16	$\pm$	0.12	&	-0.20	$\pm$	0.07	&	5061	$\pm$	18	&	3.22	$\pm$	0.04	&	1.19	$\pm$	0.02	&	-0.22	$\pm$	0.02	&	5047	$\pm$	19	&	3.20	$\pm$	0.06	&	1.19	$\pm$	0.02	&	-0.22	$\pm$	0.02	\\
22676	&	5045	$\pm$	309	&	3.01	$\pm$	0.62	&	1.13	$\pm$	0.20	&	0.12	$\pm$	0.22	&	5147	$\pm$	32	&	3.11	$\pm$	0.08	&	1.47	$\pm$	0.04	&	0.06	$\pm$	0.03	&	5109	$\pm$	36	&	3.01	$\pm$	0.09	&	1.40	$\pm$	0.04	&	0.06	$\pm$	0.03	\\
23549	&	4885	$\pm$	204	&	2.88	$\pm$	0.41	&	1.21	$\pm$	0.20	&	0.27	$\pm$	0.20	&	5038	$\pm$	33	&	2.99	$\pm$	0.07	&	1.51	$\pm$	0.03	&	0.18	$\pm$	0.03	&	4992	$\pm$	39	&	2.96	$\pm$	0.10	&	1.48	$\pm$	0.04	&	0.16	$\pm$	0.03	\\
\end{tabular}
\end{table}
\end{landscape}
\begin{landscape}
\begin{table}
\contcaption{}
\begin{tabular}{r|lccc|lccc|lccc}
\hline\hline
 & \multicolumn{4}{c|}{HM07 line-list}	& \multicolumn{4}{c|}{S08 line-list} 	& \multicolumn{4}{c}{T13 line-list}	\\	\hline		
HD	&	T$_{eff}$			&	$\log g$			&	$\xi$			&	[Fe/H]			&	T$_{eff}$			&	$\log g$			&	$\xi$			&	[Fe/H]			&	T$_{eff}$			&	$\log g$			&	$\xi$			&	[Fe/H]			\\
number	&	(K)			&	(cm s$^{-2}$)			&	(km s$^{-1}$)			&	(dex)			&	(K)			&	(cm s$^{-2}$)			&	(km s$^{-1}$)			&	(dex)			&	(K)			&	(cm s$^{-2}$)			&	(km s$^{-1}$)			&	(dex)			\\
\hline
51211	&	4860	$\pm$	71	&	2.79	$\pm$	0.13	&	1.41	$\pm$	0.10	&	-0.12	$\pm$	0.07	&	4927	$\pm$	26	&	2.73	$\pm$	0.05	&	1.50	$\pm$	0.03	&	-0.16	$\pm$	0.02	&	4899	$\pm$	27	&	2.70	$\pm$	0.08	&	1.47	$\pm$	0.03	&	-0.16	$\pm$	0.02	\\
55151	&	4792	$\pm$	62	&	2.77	$\pm$	0.14	&	1.52	$\pm$	0.08	&	-0.02	$\pm$	0.05	&	4873	$\pm$	30	&	2.83	$\pm$	0.08	&	1.51	$\pm$	0.03	&	0.01	$\pm$	0.03	&	4814	$\pm$	36	&	2.79	$\pm$	0.15	&	1.46	$\pm$	0.04	&	-0.01	$\pm$	0.02	\\
55865	&	4817	$\pm$	60	&	2.82	$\pm$	0.12	&	1.51	$\pm$	0.08	&	0.00	$\pm$	0.05	&	4926	$\pm$	38	&	2.83	$\pm$	0.07	&	1.50	$\pm$	0.04	&	0.03	$\pm$	0.03	&	4866	$\pm$	34	&	2.70	$\pm$	0.10	&	1.47	$\pm$	0.04	&	-0.01	$\pm$	0.03	\\
55964	&	4901	$\pm$	63	&	3.08	$\pm$	0.11	&	1.26	$\pm$	0.09	&	-0.10	$\pm$	0.05	&	4972	$\pm$	24	&	3.10	$\pm$	0.04	&	1.30	$\pm$	0.03	&	-0.11	$\pm$	0.02	&	4941	$\pm$	27	&	3.00	$\pm$	0.07	&	1.26	$\pm$	0.03	&	-0.12	$\pm$	0.02	\\
56478	&	4827	$\pm$	56	&	2.73	$\pm$	0.11	&	1.50	$\pm$	0.08	&	-0.10	$\pm$	0.05	&	4932	$\pm$	31	&	2.77	$\pm$	0.06	&	1.57	$\pm$	0.03	&	-0.12	$\pm$	0.03	&	4897	$\pm$	34	&	2.68	$\pm$	0.08	&	1.56	$\pm$	0.04	&	-0.15	$\pm$	0.03	\\
\end{tabular}
\end{table}
\end{landscape}
\begin{landscape}
\begin{table}
\contcaption{}
\begin{tabular}{r|lccc|lccc|lccc}
\hline\hline
 & \multicolumn{4}{c|}{HM07 line-list}	& \multicolumn{4}{c|}{S08 line-list} 	& \multicolumn{4}{c}{T13 line-list}	\\	\hline		
HD	&	T$_{eff}$			&	$\log g$			&	$\xi$			&	[Fe/H]			&	T$_{eff}$			&	$\log g$			&	$\xi$			&	[Fe/H]			&	T$_{eff}$			&	$\log g$			&	$\xi$			&	[Fe/H]			\\
number	&	(K)			&	(cm s$^{-2}$)			&	(km s$^{-1}$)			&	(dex)			&	(K)			&	(cm s$^{-2}$)			&	(km s$^{-1}$)			&	(dex)			&	(K)			&	(cm s$^{-2}$)			&	(km s$^{-1}$)			&	(dex)			\\
\hline
83380	&	4793	$\pm$	202	&	2.74	$\pm$	0.41	&	1.26	$\pm$	0.20	&	0.10	$\pm$	0.28	&	4920	$\pm$	36	&	2.88	$\pm$	0.07	&	1.47	$\pm$	0.04	&	0.06	$\pm$	0.03	&	4840	$\pm$	47	&	2.72	$\pm$	0.10	&	1.47	$\pm$	0.05	&	0.00	$\pm$	0.04	\\
83465	&	4712	$\pm$	84	&	2.58	$\pm$	0.17	&	1.49	$\pm$	0.10	&	-0.10	$\pm$	0.07	&	4839	$\pm$	31	&	2.67	$\pm$	0.06	&	1.56	$\pm$	0.03	&	-0.08	$\pm$	0.03	&	4837	$\pm$	43	&	2.72	$\pm$	0.10	&	1.52	$\pm$	0.04	&	-0.06	$\pm$	0.03	\\
84698	&	4846	$\pm$	59	&	2.71	$\pm$	0.12	&	1.37	$\pm$	0.08	&	0.01	$\pm$	0.05	&	4947	$\pm$	32	&	2.81	$\pm$	0.06	&	1.50	$\pm$	0.03	&	0.01	$\pm$	0.03	&	4870	$\pm$	36	&	2.70	$\pm$	0.08	&	1.46	$\pm$	0.04	&	-0.03	$\pm$	0.03	\\
85154	&	4948	$\pm$	204	&	2.97	$\pm$	0.41	&	1.18	$\pm$	0.20	&	0.09	$\pm$	0.57	&	5085	$\pm$	31	&	3.13	$\pm$	0.06	&	1.32	$\pm$	0.03	&	0.09	$\pm$	0.03	&	5041	$\pm$	39	&	3.06	$\pm$	0.08	&	1.34	$\pm$	0.04	&	0.07	$\pm$	0.03	\\
85250	&	5034	$\pm$	60	&	3.17	$\pm$	0.10	&	1.35	$\pm$	0.09	&	0.10	$\pm$	0.06	&	5119	$\pm$	27	&	3.16	$\pm$	0.05	&	1.35	$\pm$	0.03	&	0.09	$\pm$	0.02	&	5084	$\pm$	28	&	3.07	$\pm$	0.07	&	1.33	$\pm$	0.03	&	0.08	$\pm$	0.02	\\
\end{tabular}
\end{table}
\end{landscape}
\begin{landscape}
\begin{table}
\contcaption{}
\begin{tabular}{r|lccc|lccc|lccc}
\hline\hline
 & \multicolumn{4}{c|}{HM07 line-list}	& \multicolumn{4}{c|}{S08 line-list} 	& \multicolumn{4}{c}{T13 line-list}	\\	\hline		
HD	&	T$_{eff}$			&	$\log g$			&	$\xi$			&	[Fe/H]			&	T$_{eff}$			&	$\log g$			&	$\xi$			&	[Fe/H]			&	T$_{eff}$			&	$\log g$			&	$\xi$			&	[Fe/H]			\\
number	&	(K)			&	(cm s$^{-2}$)			&	(km s$^{-1}$)			&	(dex)			&	(K)			&	(cm s$^{-2}$)			&	(km s$^{-1}$)			&	(dex)			&	(K)			&	(cm s$^{-2}$)			&	(km s$^{-1}$)			&	(dex)			\\
\hline
129462	&	4915	$\pm$	165	&	3.00	$\pm$	0.25	&	2.04	$\pm$	0.31	&	-0.16	$\pm$	0.14	&	4985	$\pm$	27	&	2.84	$\pm$	0.04	&	1.52	$\pm$	0.03	&	-0.07	$\pm$	0.02	&	4953	$\pm$	26	&	2.75	$\pm$	0.06	&	1.48	$\pm$	0.03	&	-0.08	$\pm$	0.02	\\
129893	&	4914	$\pm$	59	&	2.91	$\pm$	0.10	&	1.43	$\pm$	0.08	&	0.01	$\pm$	0.05	&	4974	$\pm$	31	&	2.88	$\pm$	0.05	&	1.44	$\pm$	0.03	&	-0.01	$\pm$	0.03	&	4917	$\pm$	35	&	2.80	$\pm$	0.08	&	1.40	$\pm$	0.04	&	-0.03	$\pm$	0.03	\\
130650	&	4940	$\pm$	71	&	2.79	$\pm$	0.11	&	1.51	$\pm$	0.12	&	-0.07	$\pm$	0.07	&	5027	$\pm$	24	&	2.78	$\pm$	0.04	&	1.54	$\pm$	0.03	&	-0.06	$\pm$	0.02	&	5009	$\pm$	28	&	2.79	$\pm$	0.07	&	1.51	$\pm$	0.03	&	-0.06	$\pm$	0.02	\\
131376	&	5052	$\pm$	234	&	3.01	$\pm$	0.47	&	1.12	$\pm$	0.20	&	0.14	$\pm$	0.59	&	5126	$\pm$	26	&	3.11	$\pm$	0.06	&	1.43	$\pm$	0.03	&	0.08	$\pm$	0.02	&	5096	$\pm$	34	&	3.06	$\pm$	0.08	&	1.42	$\pm$	0.03	&	0.07	$\pm$	0.03	\\
132905	&	4876	$\pm$	66	&	2.54	$\pm$	0.09	&	1.46	$\pm$	0.12	&	-0.39	$\pm$	0.07	&	4954	$\pm$	22	&	2.69	$\pm$	0.05	&	1.53	$\pm$	0.02	&	-0.37	$\pm$	0.02	&	4989	$\pm$	22	&	2.61	$\pm$	0.18	&	1.55	$\pm$	0.03	&	-0.35	$\pm$	0.02	\\
\end{tabular}
\end{table}
\end{landscape}
\begin{landscape}
\begin{table}
\contcaption{}
\begin{tabular}{r|lccc|lccc|lccc}
\hline\hline
 & \multicolumn{4}{c|}{HM07 line-list}	& \multicolumn{4}{c|}{S08 line-list} 	& \multicolumn{4}{c}{T13 line-list}	\\	\hline		
HD	&	T$_{eff}$			&	$\log g$			&	$\xi$			&	[Fe/H]			&	T$_{eff}$			&	$\log g$			&	$\xi$			&	[Fe/H]			&	T$_{eff}$			&	$\log g$			&	$\xi$			&	[Fe/H]			\\
number	&	(K)			&	(cm s$^{-2}$)			&	(km s$^{-1}$)			&	(dex)			&	(K)			&	(cm s$^{-2}$)			&	(km s$^{-1}$)			&	(dex)			&	(K)			&	(cm s$^{-2}$)			&	(km s$^{-1}$)			&	(dex)			\\
\hline
175219	&	4800	$\pm$	63	&	2.56	$\pm$	0.10	&	1.49	$\pm$	0.09	&	-0.31	$\pm$	0.06	&	4896	$\pm$	25	&	2.71	$\pm$	0.05	&	1.50	$\pm$	0.03	&	-0.29	$\pm$	0.02	&	4888	$\pm$	30	&	2.64	$\pm$	0.06	&	1.52	$\pm$	0.04	&	-0.30	$\pm$	0.03	\\
175401	&	5009	$\pm$	86	&	2.98	$\pm$	0.14	&	1.79	$\pm$	0.15	&	0.04	$\pm$	0.08	&	5102	$\pm$	37	&	2.97	$\pm$	0.06	&	1.57	$\pm$	0.04	&	0.16	$\pm$	0.03	&	5042	$\pm$	42	&	2.79	$\pm$	0.12	&	1.55	$\pm$	0.04	&	0.12	$\pm$	0.04	\\
177222	&	4914	$\pm$	59	&	3.00	$\pm$	0.09	&	1.28	$\pm$	0.08	&	-0.03	$\pm$	0.06	&	4993	$\pm$	26	&	3.00	$\pm$	0.06	&	1.38	$\pm$	0.03	&	-0.06	$\pm$	0.02	&	4952	$\pm$	32	&	2.84	$\pm$	0.13	&	1.34	$\pm$	0.03	&	-0.08	$\pm$	0.03	\\
177389	&	5049	$\pm$	67	&	3.52	$\pm$	0.10	&	1.09	$\pm$	0.10	&	-0.01	$\pm$	0.05	&	5102	$\pm$	24	&	3.50	$\pm$	0.04	&	1.08	$\pm$	0.03	&	-0.03	$\pm$	0.02	&	5061	$\pm$	26	&	3.49	$\pm$	0.09	&	1.05	$\pm$	0.03	&	-0.05	$\pm$	0.02	\\
179433	&	5291	$\pm$	220	&	3.49	$\pm$	0.44	&	1.00	$\pm$	0.20	&	0.28	$\pm$	0.57	&	5168	$\pm$	22	&	3.15	$\pm$	0.04	&	1.38	$\pm$	0.02	&	-0.02	$\pm$	0.02	&	5145	$\pm$	31	&	3.16	$\pm$	0.08	&	1.39	$\pm$	0.04	&	-0.04	$\pm$	0.03	\\
\hline
\end{tabular}

\vspace{0.01cm}
{$^\dag$ For this star the parameters derived with the SO08 line-list may also be adopted as final. See Sect.~\ref{Sec3} for details.}
\end{table}
\end{landscape}


\begin{table*}
\begin{minipage}{126mm}
\caption{Stellar parameters of 74 stars common with our sample. References are provided into the table.\label{tbl2}}
\begin{tabular}{cccccccccc}
\hline\hline
 HD	&	T$_{eff}$	&$\epsilon$(T$_{eff})$		&	$\log g$	&	$\epsilon(\log g)$			&	[Fe/H] & $\epsilon$([Fe/H]) &	$\xi$	&$\epsilon(\xi)$ & Ref.	\\
 number	&	(K)			&   & 	(cm s$^{-2}$)		&	&	(dex)			&	 &	(km s$^{-1}$)			&		&		\\
\hline
496	&	4893	&	--	&	3	&	--	&	0.13	&	--	&	1.5	&	--	&	(1)	\\
770	&	4696	&	100	&	2.68	&	0.1	&	-0.12	&	0.1	&	1.4	&	0.2	&	(2)	\\
1737	&	5020	&	--	&	2.73	&	--	&	0.16	&	0.12	&	1.44	&	--	&	(3)	\\
5457	&	4780	&	--	&	2.74	&	--	&	-0.07	&	0.14	&	1.38	&	--	&	(3)	\\
8651	&	4708	&	100	&	2.67	&	0.1	&	-0.18	&	0.1	&	1.4	&	0.2	&	(2)	\\
9362	&	4754	&	100	&	2.61	&	0.1	&	-0.28	&	0.1	&	1.3	&	0.2	&	(2)	\\
\end{tabular}
\end{minipage}
\end{table*}

\begin{table*}
\begin{minipage}{126mm}
\contcaption{}
\begin{tabular}{cccccccccc}
\hline\hline
 HD	&	T$_{eff}$	&$\epsilon$(T$_{eff})$		&	$\log g$	&	$\epsilon(\log g)$			&	[Fe/H] & $\epsilon$([Fe/H]) &	$\xi$	&$\epsilon(\xi)$ & Ref.	\\
 number	&	(K)			&   & 	(cm s$^{-2}$)		&	&	(dex)			&	 &	(km s$^{-1}$)			&		&		\\
\hline
213009	&	4800	&	--	&	2	&	--	&	-0.2	&	--	&	1.3	&	--	&	(13)	\\
215104	&	4737	&	100	&	2.62	&	0.1	&	-0.2	&	0.1	&	1.5	&	0.2	&	(2)	\\
216763	&	4841	&	100	&	2.72	&	0.1	&	-0.21	&	0.1	&	1.4	&	0.2	&	(2)	\\
220790	&	4850	&	--	&	2.67	&	--	&	-0.27	&	0.13	&	1.37	&	--	&	(3)	\\
222433	&	4860	&	--	&	3.05	&	--	&	-0.3	&	--	&	2.1	&	--	&	(7)	\\
166599	&	5005	&	50	&	2.6	&	0.25	&	-0.03	&	0.09	&	--	&	--	&	(11)	\\
175219	&	4720	&	--	&	2.44	&	--	&	-0.32	&	--	&	--	&	--	&	(12)	\\
24160	&	5010	&	--	&	--	&	--	&	--	&	--	&	--	&	--	&	(5)	\\
94890	&	4802	&	--	&	--	&	--	&	--	&	--	&	--	&	--	&	(5)	\\
96566	&	4901	&	--	&	--	&	--	&	--	&	--	&	--	&	--	&	(5)	\\
116243	&	5181	&	--	&	--	&	--	&	--	&	--	&	--	&	--	&	(5)	\\
134505	&	4990	&	--	&	--	&	--	&	--	&	--	&	--	&	--	&	(5)	\\
\hline
\end{tabular}
{			
(1) 	\citet{1986A&A...169..201G};
(2)  	\citet{2007MNRAS.382..553L};
(3)  	\citet{2011A&A...536A..71J};
(4)	\citet{2006A&A...458..609D};
(5)	\citet{1998A&A...339..858D};
(6)	\citet{2007A&A...475.1003H};
(7)	\citet{1990ApJS...74.1075M};
(8)	\citet{2004A&A...425..187T};
(9)	\citet{1992MNRAS.256..535J};
(10)	\citet{1991ApJS...75..579L};
(11)	\citet{1999A&A...348..487R};
(12)	\citet{2008A&A...484L..21M};
(13)	\citet{1981A&A....93..315F}
}
\label{lastpage}
\end{minipage}
\end{table*}
\end{document}